\documentclass[12pt,preprint]{aastex}

\newcommand{\myemail}{shogo@kusastro.kyoto-u.ac.jp}
\shorttitle{GC Polarization 2}
\shortauthors{Nishiyama et al.}

\begin{document}

\title{Magnetic Field Configuration at the Galactic Center
Investigated by Wide Field Near-Infrared Polarimetry:\\
Transition from a Toroidal to a Poloidal Magnetic Field}

\author{Shogo Nishiyama\altaffilmark{1},
Hirofumi Hatano\altaffilmark{2},
Motohide Tamura\altaffilmark{3},
Noriyuki Matsunaga\altaffilmark{4},
Tatsuhito Yoshikawa\altaffilmark{1},
Takuya Suenaga\altaffilmark{5},
James H. Hough\altaffilmark{6},
Koji Sugitani\altaffilmark{7},
Takahiro Nagayama\altaffilmark{2},
Daisuke Kato\altaffilmark{8},
and Tetsuya Nagata\altaffilmark{1}
}

\altaffiltext{1}{Department of Astronomy, Kyoto University, 
Kitashirakawa-oiwake-cho, Sakyo-ku, Kyoto 606-8502, Japan; \myemail}

\altaffiltext{2}{Department of Astrophysics, Nagoya University, 
Furo-cho, Chikusa-ku, Nagoya 464-8602, Japan}

\altaffiltext{3}{National Astronomical Observatory of Japan, 
2-21-1, Osawa, Mitaka, Tokyo 181-8588, Japan}

\altaffiltext{4}{Kiso Observatory, Institute of Astronomy, University of Tokyo,
10762-30, Mitake, Kiso, Nagano 397-0101, Japan}

\altaffiltext{5}{Department of Astronomical Sciences, Graduate University for Advanced Studies (Sokendai),
2-21-1, Osawa, Mitaka, Tokyo 181-8588, Japan}

\altaffiltext{6}{Centre for Astrophysics Research, Science and Technology Research Institute,
University of Hertfordshire, Hatfield AL10 9AB, UK}

\altaffiltext{7}{Graduate School of Natural Sciences, Nagoya City University,
1, Yamanohata, Mizuho-cho, Mizuho-ku, Nagoya 467-8501, Japan}

\altaffiltext{8}{Institute of Space and Astronomical Science, Japan Aerospace Exploration Agency,
3-1-1, Yoshinodai, Chuo-ku, Sagamihara, Kanagawa, 252-5210, Japan}


\begin{abstract}

We present a large-scale view of the magnetic field
in the central $2\degr \times 2\degr$ region of our Galaxy. 
The polarization of point sources has been measured 
in the $J$, $H$, and $K_S$ bands
using the near-infrared polarimetric camera SIRPOL on the 1.4 m telescope IRSF.
Comparing the Stokes parameters between high extinction stars 
and relatively low extinction ones,
we obtain polarization originating from magnetically aligned dust grains
in the central few-hundred pc of our Galaxy.
We find that near the Galactic plane, the magnetic field is almost parallel to the Galactic plane
(i.e., toroidal configuration)
but at high Galactic latitudes ($\mid b \mid > 0\fdg4$), 
the field is nearly perpendicular to the plane (i.e., poloidal configuration). 
This is the first detection of a smooth transition of 
the large-scale magnetic field configuration in this region.

\end{abstract}


\section{INTRODUCTION}
\label{sec:Intro}

It has long been a subject of controversy whether the magnetic field (MF)
in the central region of our Galaxy has a dipolar or toroidal geometry on a large scale.
Previous observations have been mainly based on 
sub-millimeter polarimetric observations of dense molecular clouds,
or radio observations of non-thermal radio filaments. 
The former show that the MFs in the dense molecular clouds 
run almost parallel to the Galactic plane, indicating a {\it toroidal} MF \citep{Novak03,Chuss03}.
By contrast, 
the radio filaments are aligned nearly perpendicular to the Galactic plane \citep{Yusef84,Yusef04,LaRosa04},
and are highly polarized along the filaments' long axes \citep{Tsuboi86,Yusef87},
suggesting a large-scale, pervasive {\it poloidal} MF in the intercloud medium.

The MFs both in dense molecular clouds and in the diffuse interstellar (intercloud) medium
are observable via optical and near-infrared (NIR) polarimetry of intrinsically unpolarized stars 
with the polarization attributed to differential extinction by dust grains 
aligned with their short axis along the direction of the MF \citep[e.g.,][]{Mathewson70}.
The polarization vectors of such stars trace
the orientation of the MF projected onto the plane of the sky,
and measurements of stars at different distances reveal
the change of the projected MF orientation along the line of sight.
Although there are numerous stars near the Galactic center (GC),
high extinction by the intervening dust grains means that
infrared polarimetry is ideal \citep[e.g.,][]{Tamura87} in understanding the MF configuration in the GC.
However, infrared polarimety toward the GC has been carried out for 
only a limited number of stars so far \citep[e.g.,][]{Hough78,Kob83,Bailey84,Eckart95,Ott99}.

In this letter, we report observations of the GC region
with a wide-field NIR polarimeter.
In the first paper of this series \citep{Nishi09}, we presented the MF configuration
in the $20\arcmin \times 20\arcmin$ region of the GC.
In the following observations from 2007,
more than 300,000 stars are detected 
in the central $2\degr \times 2\degr$ region
with an error in the degree of polarization less than 1\%.
The statistical treatment of the large number of stars
enables us to obtain the first large-scale view of the MF in the GC.

\section{OBSERVATIONS AND DATA REDUCTION}
\label{sec:Obs}

Polarimetry for a large number of stars 
in the central $2\degr \times 2\degr$ region of the Galaxy
was obtained with IRSF/SIRPOL between July 2006 and August 2009.
SIRPOL consists of a single-beam polarimeter 
\citep[a half-wave plate rotator unit and a fixed wire-grid polarizer;][]{Kandori06}
and the NIR imaging camera SIRIUS
\citep[Simultaneous Infrared Imager for Unbiased Survey;][]{Nagas99,Nagay03},
and is attached to the 1.4-m telescope IRSF (Infrared Survey Facility).
The detectors are three 1024 $\times$ 1024 HgCdTe arrays,
with a scale of 0\farcs45 pixel$^{-1}$.  
SIRPOL provides images of a 7\farcm7 $\times$ 7\farcm7 area of sky 
in three NIR bands, $J$ ($1.25\mu$m), $H$ ($1.63\mu$m),
and $K_S$ ($2.14\mu$m), simultaneously.
The filter system of IRSF/SIRPOL is similar to 
the Mauna Kea Observatory system \citep{Tokunaga00}.

10-s exposures were made, at each 4 wave plate angles
($0\fdg0$, $22\fdg5$, $45\fdg0$, $67\fdg5$), and
at each of 10 dithered positions.
The weather condition was photometric, with a mean seeing of 
$\sim$1\farcs3 ($J$), $\sim$1\farcs2 ($H$), and $\sim$1\farcs1 ($K_S$).
Twilight flats were obtained before and after the observations.
The IRAF (Image Reduction and Analysis Facility)\footnote{
IRAF is distributed by the National Optical Astronomy
Observatory, which is operated by the Association of Universities for
Research in Astronomy, Inc., under cooperative agreement with
the National Science Foundation.}
software package was used
to perform dark- and flat-field corrections,
followed by sky background estimation and subtraction.

\section{ANALYSIS}
\label{sec:Ana}

To obtain the photometric magnitudes and errors,
we used the DAOPHOT package \citep{Stetson87}.
The sources detected in Stokes $I$ images 
[$I=(I_{0\degr}+I_{22\fdg5}+I_{45\degr}+I_{67\fdg5})/2$]
were input to the ALLSTAR task for PSF-fitting photometry.
About 10 sources were used to construct the PSF in each image.
Photometric calibration was performed with reference to 
the magnitudes of point sources 
in the 2MASS catalog \citep{Skrutskie06}.

The Stokes parameters $I$, $Q$, and $U$ for point sources
were determined from aperture polarimetry,
which gives a better photometric result than PSF-fitting photometry in our GC data.
We obtained the intensity for each wave plate angle
($I_{0\degr}$, $I_{22\fdg5}$, $I_{45\degr}$, $I_{67\fdg5}$)
with IRAF DAOFIND and APPHOT tasks.
The size of the aperture is slightly different 
among fields due to variations of the seeing,
and an aperture size of 1.5 $\times$ FWHM was adopted,
which was measured on Stokes $I$ images.
Our data analysis results in $\sim 140,000 (J)$, $\sim 360,000 (H)$, 
and $\sim 380,000 (K_S)$ sources in the observed area
with an error $\delta P$ in the degree of polarization $P$ 
less than 1\%, where
\[ P = \sqrt{(Q^2+U^2)}/I,\]
and
\[ \delta P = \frac{1}{P} \sqrt{ \left( \frac{Q}{I} \right)^2 \left[ \delta \left( \frac{Q}{I} \right) \right]^2 +
\left( \frac{U}{I} \right)^2 \left[ \delta \left( \frac{U}{I} \right) \right]^2 }.\]
Polarimetric accuracy was checked with duplicate sources
in overlapping regions of adjacent fields 
observed under different atmospheric condition.
The means of the average differences of $Q/I$ and $U/I$ were
0.26 \% and 0.27 \% in the $H$ band, and
0.25 \% and 0.26 \% in the $K_S$ band, respectively.

To derive the MF configuration at the GC,
we separated stars into two subgroups according to their colors:
``blue'' and ``red'' stars.
Most of the stars detected in the observations are late-type giants near the GC,
and their intrinsic $H-K_S$ colors are almost the same\footnote{
The standard deviation of the intrinsic $H-K_S$ colors of the observed giants in the GC
is calculated to be 0.06 from the Galactic model by \citet{Wainscoat92}.
This is small enough compared to the mean of the color ranges of red and blue stars, 
$2 \sigma_{H-K_S} = 0.38$.
Therefore we proceed on the supposition that the observed colors 
of the giants depend on the distance to them.}.
So their observed colors depend on the amount of the interstellar extinction
which in turn depends generally on the distance to the stars.
The ``red'' (``blue'') stars suffer strong (weak) extinction, 
and thus they are distributed at the far (near) side of the GC.
The light from stars at the near side of the GC (i.e., ``blue'' stars) 
is transmitted through the interstellar dust in the Galactic disk,
and the polarization is produced there, whereas
the light from stars at the far side of the GC (i.e., ``red'' stars) 
is transmitted through the dust in the disk {\it and} the GC. 
By comparing the Stokes parameters ($Q/I$ and $U/I$) of the ``blue'' and ``red'' stars,
we obtain polarization of the GC component 
\citep[see][for more detail]{Nishi09}.

As a first step in deriving the MF configuration at the GC, 
the field was divided into sub-fields of $2\arcmin \times 2\arcmin$,
and $H-K_S$ histograms were made for each sub-field.
These histograms were fitted with the Gaussian function, and 
means ($m_{H-K_S}$) and sigmas ($\sigma_{H-K_S}$) 
were calculated for each sub-field.
Using the mean and sigma values, 
the stars were divided into two subgroups:
``blue'' stars with the color of $(m_{H-K_S} - 2 \sigma_{H-K_S}) < (H-K_S) < m_{H-K_S}$,
and ``red'' stars with $m_{H-K_S} < (H-K_S) < (m_{H-K_S} + 2 \sigma_{H-K_S})$.

Mean ($m_{H-K_S}$) and sigma ($\sigma_{H-K_S}$) in the $H-K_S$ histograms
were determined for stars with $H < 15.5$, which are bright enough
to avoid the influence of the limiting magnitude, $H \approx 16.6$
\citep[see also Fig. 7 of][]{Nishi09}.
This means that our observations are equally sensitive to stars
on the far and near sides of the GC, and 
using this criterion, stars on both sides of the GC can be distinguished.
The ``blue'' and ``red'' stars are in fact distributed evenly 
in the $2\arcmin \times 2\arcmin$ sub-fields.
Hence, it can be safely assumed that
the peaks in the $H-K_S$ histograms correspond
to a real peak of spatial distribution of stars along the line of sight.

Next, $Q/I$ and $U/I$ histograms in the $K_S$ band
were constructed for the ``blue'' and ``red'' stars 
with $\delta P < 3 \%$ in each sub-field.
Their means, 
$<Q/I>_{\mathrm B}$, $<Q/I>_{\mathrm R}$,
$<U/I>_{\mathrm B}$, and $<U/I>_{\mathrm R}$,
were used to obtain the degree $P$ and position angle $\theta$ 
of polarization for ``red'' minus ``blue'' (i.e., GC component) 
with the following equations \citep{Goodrich86}:
\begin{eqnarray}
P_{\mathrm {R-B}} = \sqrt{ \left( \left< \frac{Q}{I} \right>_{\mathrm R} - \left< \frac{Q}{I} \right>_{\mathrm B} \right)^{2} 
  + \left( \left< \frac{U}{I} \right>_{\mathrm R} - \left< \frac{U}{I} \right>_{\mathrm B}  \right)^{2} }, 
\label{eq:Prb}
\end{eqnarray}
\begin{eqnarray}
\theta_{\mathrm {R-B}} = \frac{1}{2} \arctan \left( 
\frac{\left< \frac{U}{I} \right>_{\mathrm R} - \left< \frac{U}{I} \right>_{\mathrm B}}
     {\left< \frac{Q}{I} \right>_{\mathrm R} - \left< \frac{Q}{I} \right>_{\mathrm B}} \right).
\label{eq:Thetarb}
\end{eqnarray}
The errors of $<Q/I>$ and $<U/I>$ were calculated from the standard error on the mean
$\sigma/\sqrt{N}$ of the $Q/I$ and $U/I$ histograms,
where $\sigma$ is the standard deviation and $N$ is the number of stars.
A few sub-fields with strong extinction are excluded from our analysis.
We confirmed that the ratio of the number of the ``red'' and ``blue'' stars 
in each sub-field has a mean of $\approx 1$
even near the Galactic plane ($l < 0\fdg2$),
suggesting no bias effect to detect more stars at the near side
than the far side of the GC due to strong extinction.
Finally, 
$<Q/I>$ and $<U/I>$ are averaged in a circle of 2\farcm4 radius with a 3\farcm0 grid,
and $P_{\mathrm {R-B}}$ and $\theta_{\mathrm {R-B}}$ are calculated 
with the equations (\ref{eq:Prb}) and (\ref{eq:Thetarb}).
The resultant mean and rms of $P_{\mathrm {R-B}}$ are 
0.29 \% and 0.12 \% in the $K_S$ band, respectively.

The strong concentration of stars towards the GC
allows measurement of the MF structure in the GC region.
Most of the stars detected in our observations have an intrinsic color 
of red giants with spectral type of K and M.
We used a Galactic model \citep{Wainscoat92} 
to show the degree of central concentration of the stars along the line of sight,
and found that more than 85 \% of stars within the central $2\degr \times 2\degr$ region
are expected to be within 300 pc of the GC.
So the observed 
position angles of polarization of the GC component show those of MFs, 
averaged along the lines of sight, in the central few-hundred pc.

\section{RESULTS AND DISCUSSION}

\begin{figure}[h]
  \begin{center}
    \epsscale{1.0}
    \plotone{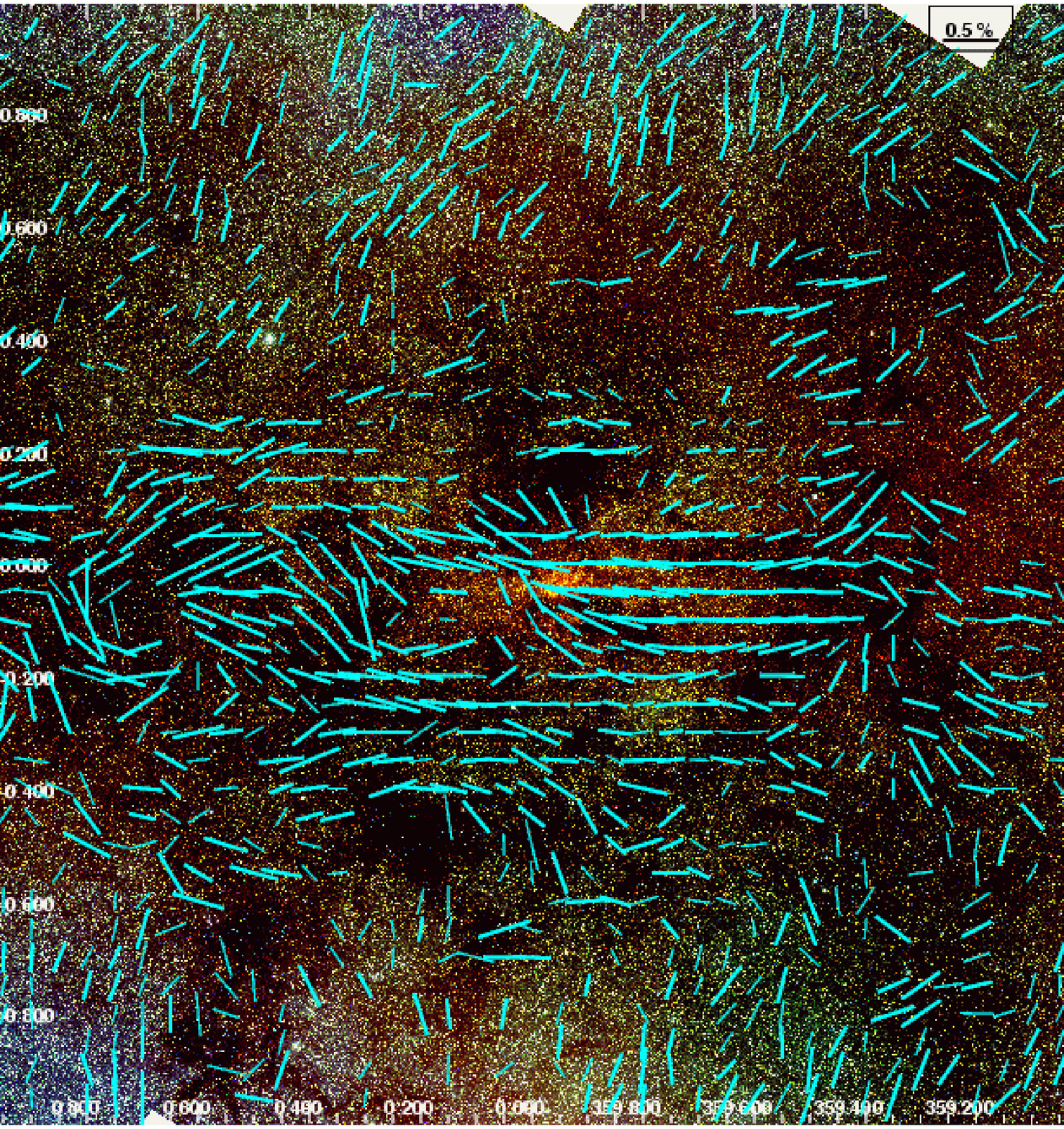}
    \caption{
      NIR mosaic image of the Galactic center region 
      covering $2\fdg0 \times 2\fdg0$ in the Galactic coordinate,
      taken with the IRSF telescope and NIR camera SIRIUS.
      The three NIR bands are $J$ (blue, $1.25\mu$m), 
      $H$ (green, $1.63\mu$m), and $K_S$ (red, $2.14\mu$m).
      The central-parsec star cluster of the Galactic center
      is the bright yellow blob in the center of the image.
      Observed $E$-vectors of polarization for the Galactic center components 
      in the $K_S$ band are also plotted 
      with cyan bars whose length indicates the degree of polarization.
      The vectors are averaged in a circle of 2\farcm4 radius with a 3\farcm0 grid,
      and plotted with thick bars (detected with more than $3 \sigma$)
      and thin bars (detected with $2-3 \sigma$).
    }
  \label{fig:Vmap}
 \end{center}
\end{figure}

\begin{figure}[h]
 \begin{center}
   \epsscale{.80}
   \plotone{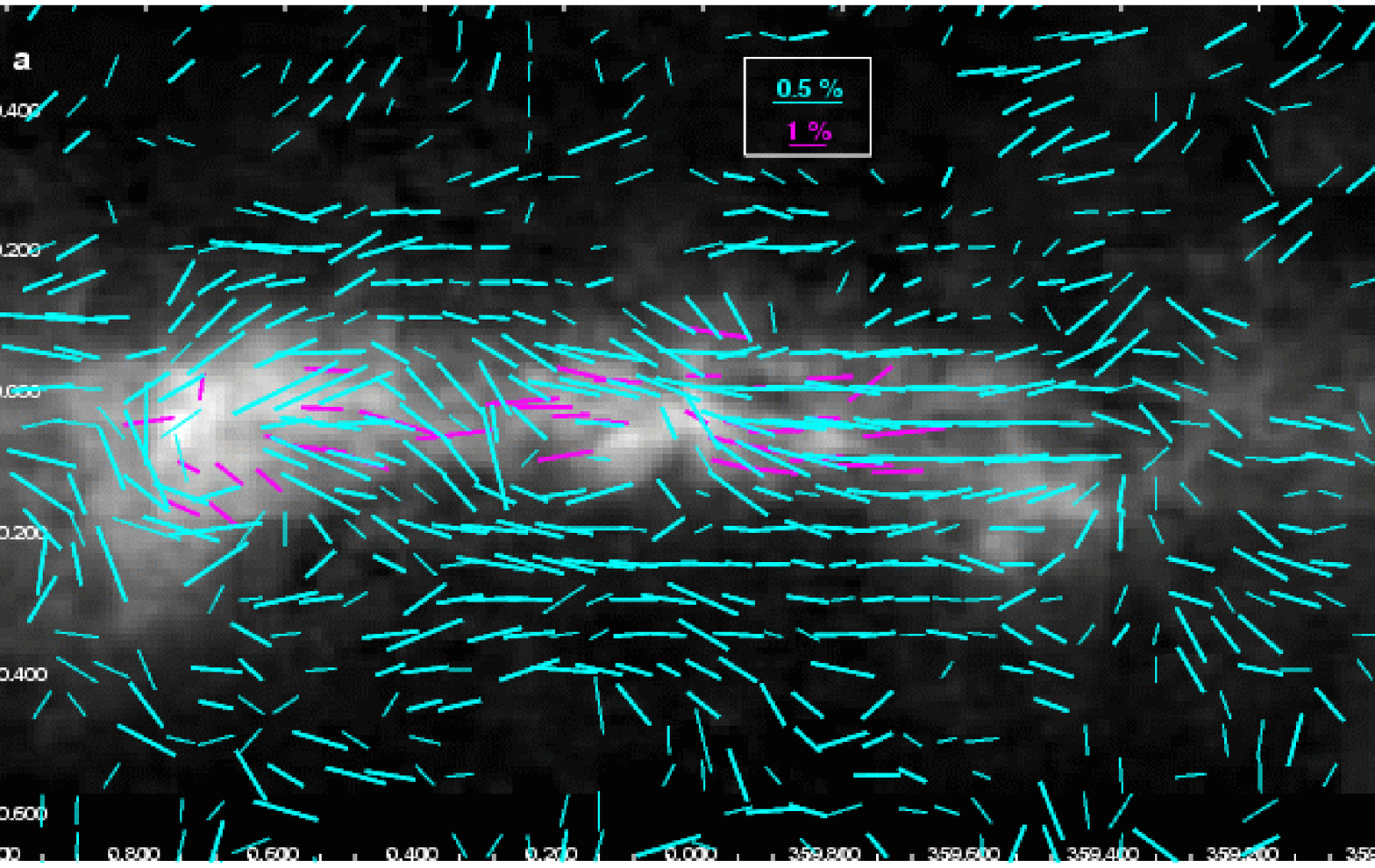}
   \plotone{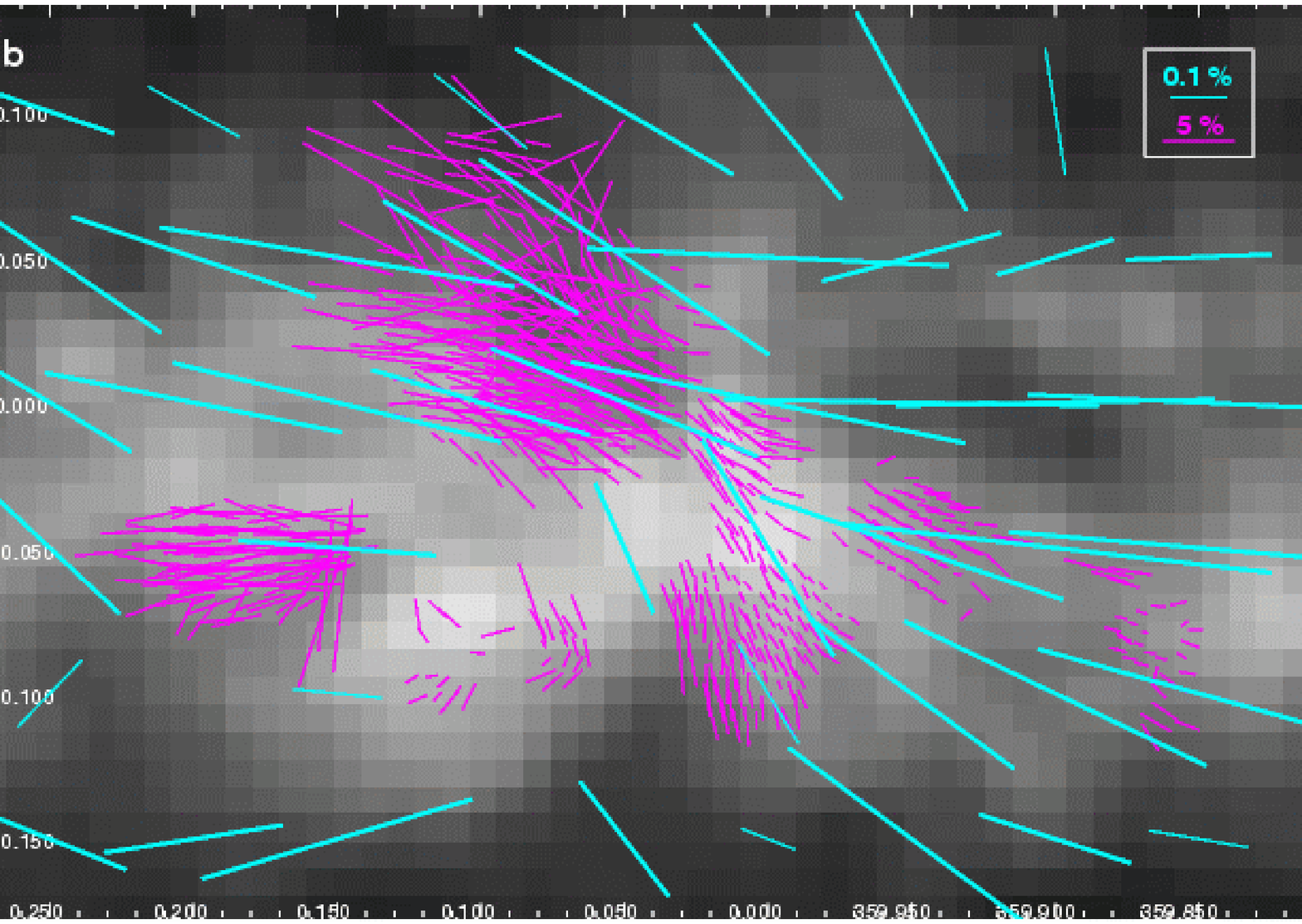}
    \caption{
      Maps of the $^{12}$CO $J$=1-0 line emission integrated over the velocity range
      $V_{\mathrm{LSR}} = -100$ to $+100$ km s$^{-1}$ in Galactic coordinate \citep{Oka98}.
      Pink bars in the panel (a) are drawn parallel to 
      the magnetic field direction derived from polarimetry at 450 $\mu$m \citep{Novak03},
      and those in the panel (b) from measurements at
      60 $\mu$m, 100 $\mu$m \citep{Dotson00}, and 350 $\mu$m \citep{Novak00}.
      The MF directions measured at near-infrared wavelength
      (same as Fig. \ref{fig:Vmap}) are also plotted
      with cyan bars whose length indicates the degree of polarization.
      Note that since emissive polarization in the sub-mm/FIR wavelengths is
      perpendicular to the magnetic field, vectors orthogonal to it are shown here. 
    }
  \label{fig:VmapSubmmCO}
 \end{center}
\end{figure}

Fig. \ref{fig:Vmap} shows the
MF direction ($E$-vectors of polarization of the GC component in the $K_S$ band) in the GC.
This is the first large-scale view of the MF in this region.
The MF directions in the dense molecular clouds, measured
with sub-millimeter (sub-mm) and far-infrared (FIR) polarimetry,
are overplotted on the $K_S$ data in Fig. \ref{fig:VmapSubmmCO}.
The panel (a) shows
the MF directions derived from polarimetry in a large survey at 450 $\mu$m \citep{Novak03}.
Near the Sagittarius A region, sub-mm polarimetry has been carried out with better spatial resolution
\citep[Fig. \ref{fig:VmapSubmmCO} b;][]{Dotson00,Novak00,Chuss03}.
In both cases, the MF directions derived from the sub-mm/FIR polarimetry are well 
matched with the NIR observations.
It follows that the position angle of polarization derived from NIR polarimetry
can be used to determine the direction of the MF in the GC.

\begin{figure}[h]
 \begin{center}
  \epsscale{.40}
    \plotone{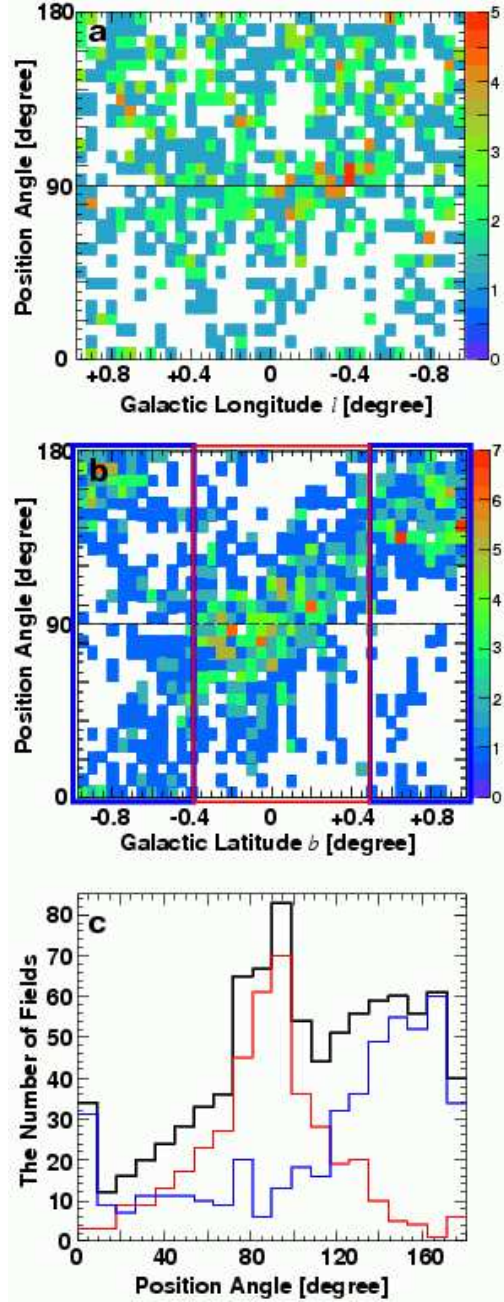}
    \caption{
      Directions of the magnetic fields in the Galactic center
      as a function of the Galactic longitude (a) and latitude (b).
      The number of field per bin is shown in different colors,
      as indicated in the color scale.
      The MF direction is defined as the angular offset in degrees
      from north through east in Galactic coordinates.
      The magnetic field swings from the direction along the Galactic plane at $b \sim 0\degr$
      to the direction nearly perpendicular to the plane at $\mid b \mid \sim 1\fdg0$.
      Histograms of the magnetic field direction
      at $\mid b \mid \leq 0\fdg4 $ (red),
      at $\mid b \mid > 0\fdg4 $ (blue), and both (black) are 
      also shown in the panel (c).
      The red histogram has a clear peak at the direction 
      almost parallel to the Galactic plane,
      while the blue one has a peak at $\sim 170\degr$,
      almost perpendicular to the plane. 
    }
  \label{fig:Hists}
 \end{center}
\end{figure}

The MF direction has strong dependence on the Galactic latitude, 
i.e., distance from the Galactic plane,
but not on the Galactic longitude (Fig. \ref{fig:Hists}).
The MF direction (i.e., the position angle of polarization) 
is defined as the angular offset in degrees
from north through east in Galactic coordinates.
The histogram of the MF direction at $\mid b \mid \leq 0\fdg4$ has 
a clear peak of $\sim 90\degr$, 
suggesting a toroidal MF configuration in this region.
At higher Galactic latitude ($\mid b \mid >  0\fdg4$), 
the mean MF direction appears to swing from $90\degr$ to $\sim 170\degr$
which is nearly perpendicular to the Galactic plane.
These results suggest a transition of the large-scale MF configuration 
from toroidal to poloidal in the GC region, at $\mid b \mid \sim 0\fdg4$.

In the central few hundred pc region,
there are some structures surrounding the GC, such as 
the 180-pc (or expanding) molecular ring \citep{Kaifu72,Scoville72}, 
and the GC molecular arms composing a ring of radius of 
$\sim 120$ pc \citep{Sofue95}.
It is therefore difficult to completely exclude the possibility that
we are preferentially probing magnetic fields in these molecular structures.
However, it seems to be convincing that we have detected the magnetic fields 
in molecular clouds much closer to the GC, from the following points.
As shown in Fig. \ref{fig:VmapSubmmCO} \citep[see also Fig. 10 of][]{Nishi09},
we are able to replicate the detailed structure of 
the sub-mm/FIR magnetic field maps in the circumnuclear disk, and
two molecular clouds M$-0.02-0.07$ (+50 km s$^{-1}$ cloud)
and M$-0.13-0.08$ (+20 km s$^{-1}$ cloud).
These two molecular clouds are located within several tens of pc from the GC,
and the possibility of contribution of foreground emission was 
also ruled out \citep{Novak00}.
Hence we can probe the magnetic fields very close to the GC.

Previously, the MF configuration of the GC has been viewed as
poloidal in the diffuse, interstellar (intercloud) medium,
and approximately parallel to the Galactic plane
only in the dense molecular clouds \citep[e.g.,][]{Ferriere09}.
The new data presented here shows a toroidal MF
prevails at $\mid b \mid < 0\fdg4$, even outside the dense molecular clouds
(Fig. \ref{fig:VmapSubmmCO}), which
radio \citep{Oka98} and sub-mm \citep{Pierce-Price00} surveys show are
mostly confined to within $\sim 0\fdg2$ of the Galactic plane.
Toward higher Galactic latitudes ($\mid b \mid > 0\fdg4$), 
the field changes from a toroidal to a poloidal configuration.

Since SIRPOL also provides images in the $H$ band,
it is possible to compare these with the $K_S$-band results (Fig. \ref{fig:Vmap}).
Most of the position angles of polarization in the $H$ band
show the same direction as that in the $K_S$ bands within their errors. 
In addition, there is a good correlation between the degree of polarization $P$
in the $H$ and $K_S$ bands with $P_{K_S} = 0.611 \times P_{H}$.
This corresponds to $\beta = 1.85$ for a power-law approximation
$P_{\lambda} \propto \lambda^{-\beta}$
which represents the wavelength dependence of interstellar polarization
at NIR wavelengths, consistent with previous
results toward the GC \citep[$\beta \approx 1.8-2.0$;][]{Nagata94,Hatano10}.
The correlation coefficients in the degree and position angle of polarization
are 0.71 and 0.92, respectively, 
showing strong correlation between the $H$ and $K_S$ bands.
These results strongly suggest that
the differential extinction by magnetically aligned dust grains 
is responsible for the observed NIR polarization.

The existence of the toroidal MF
even in the intercloud medium near the Galactic plane is consistent with 
the observations of Galactic center diffuse X-ray (GCDX) emission \citep{Koyama89}.
The spatial distribution of the 6.7 keV line emission from Helium-like ions of iron
is elliptical with a major axis approximately along the Galactic plane \citep{Yamauchi90}.
Although the origin of the GCDX is as yet unclear,
recent observations with the Suzaku satellite found
a smooth and monotonic variation of the ionization temperature along the Galactic longitude,
and a clear difference of distribution between the 6.7 keV line flux and 
an integrated point source flux,
suggesting hot, diffuse interstellar plasma in origin \citep{Koyama07}.
If a poloidal MF is pervasive in this region,
since the gravitational well of the GC region is far too shallow to confine them,
the hot plasma must be rushing out of the plane vertically as a galactic wind.
Such a vertical extension of the GCDX has not been observed.

The large-scale MF configuration has important implications
for understanding the origin of a variety of magnetic phenomena.
The double helix nebula \citep{Morris06} 
extends almost perpendicular to the Galactic plane from $b=0\fdg5$ to $0\fdg8$.
It exists in the region of the large-scale poloidal MF,
so the nebula probably results from a dynamically ordered interstellar phenomenon 
involving its local interstellar MFs.
On the other hand, a large part of the non-thermal radio filaments
exist at $\mid b \mid \la 0\fdg4$ where the toroidal field is dominant \citep{Yusef04,LaRosa04}. 
Hence the filaments do not run along the MF directions we obtained;
rather, they are almost perpendicular to the local MF.
This implies that it is important to reconsider
the simple picture in which
the non-thermal filaments are illuminated flux tubes
within a uniform poloidal MF \citep{Morris90}, 
for the origin of the filaments.

\acknowledgements

We are grateful to T. Oka for kindly providing the $^{12}$CO $J$=1-0 data. 
We thank the staff of the South African Astronomical Observatory (SAAO)
for their support during our observations.
S. N. and H. H. are financially supported by the Japan Society for the Promotion of Science (JSPS) 
through the JSPS Research Fellowship for Young Scientists.
This work was partly supported by
the Grants-in-Aid for Young Scientists (B) 19740111, 
Scientific Research (C) 21540240, 
and the Global COE Program 
"The Next Generation of Physics, Spun from Universality and Emergence" 
from the Ministry of Education, Culture,
Sports, Science and Technology (MEXT) of Japan.
M. T. has been supported by the MEXT, Grants-in-Aid 19204018 and 22000005.
This publication makes use of data from the Two Micron All Sky Survey, 
a joint project of the University of Massachusetts,
the Infrared Processing and Analysis Center, 
the National Aeronautics and Space Administration, 
and the National Science Foundation.



\end{document}